# Study of Propagating Modes and Reflectivity in Bragg Filters with AlxGa1-xN/GaN Material Composition

Sourangsu Banerji
*Department of Electronics & Communication Engineering,*
*RCC Institute of Information Technology, India*
*sourangsu.banerji@gmail.com*

*Abstract*

*In this paper, forward and backward propagating waves and reflectivity in an optical waveguide structure namely the fiber Bragg reflector also considered as a one dimensional photonic crystal, are analytically computed using coupled mode theory for different grating lengths and coupling conditions. AlxGa1-xN/GaN material composition is considered as unit block of the periodic organization, and refractive index of AlxGa1-xN is taken to be dependent on material composition, bandgap and operating wavelength following Adachi's model. The structure being considered is the Bragg grating where increase in grating length enhances the reflection of electromagnetic wave, and strong coupling provides larger bandgap spectral width. Input wavelength is made different from Bragg wavelength to study the characteristics of propagating waves. A suitable combination of grating length and coupling coefficient is helpful in designing the photonic bandgap at 1550 nm wavelength. These characteristic curves can be utilized to study how waves propagate through the optical waveguides which have a special place in optical communications*

**Keywords:** *Optical Waveguides, Forward and Backward Waves, Bragg Wavelength, Reflectivity, Coupling Coefficient*

## 1. Introduction

Photonic crystal is a periodic arrangement of dielectric materials [1] where localization of electromagnetic wave propagating inside is controlled by suitably varying the structural parameters. This is possible due to the formation of photonic bandgap, and the property can be used in photonic integrated circuit [2], optical transmitter [3], and optical receiver [4], and photonic crystal fiber [5], quantum information processing [6]. This novel microstructure already replaced conventional optical fiber for efficient communication.

Various types of confinements has already been considered by theoretical researchers for analysis of photonic crystal, but it is suggested that only 1D and 2D structures are efficient enough to fabricate and also implementation purpose for different optoelectronic integrated circuits. Among them, 1D structure is very convenient to study because of ease of mathematical modeling. Propagating wave analysis in 1D structure is useful for designing four-wave mixing analysis in nonlinear photonic crystal [7]. Suitable dielectric materials are used to characterize modal dispersion in 1D crystal [8]. Incorporation of semiconductor nanostructure makes it more interesting when filter characteristics is considered [9] including the effect of polarization of incident light.

The present paper deals with propagation wave characteristics of forward and backward waves as a function of grating length in an optical waveguide structure when Bragg wavelength is set at 1550 nm. Coupled mode theory is used to solve the problem using Bragg condition. Refractive indices of the materials are considered as function of material composition (x), bandgap and operating wavelength. Reflectivity is computed from the knowledge of wave propagation for different coupling conditions and also for variable dimensions. The spectral width of the structure i.e., photonic bandgap is measured from the reflection coefficient analysis.

Use of AlxGa1-xN/GaN material composition was taken up because as shown in [18] that the effect of carrier localization in undoped AlGaN alloys enhances with the increase in Al contents and is related to the insulating nature of AlGaN of high Al contents. The use of high Al-content AlGaN layer is also expected to increase the overall figure of merit of the AlGaN/GaN due to the combined advantages of enhanced band offset, lattice mismatch induced piezoelectric effect. Thus improving the material quality of high Al content AlGaN alloys is also of crucial importance for fabricating high performance AlGaN/GaN structures. The advantages of GaN can be summarized as ruggedness, power handling and low loss [17], [19].

## 2. Mathematical Modeling

### 2.1. Coupled Optical Modes

Propagating modes on optical waveguides can interact, and therefore be coupled, just as the mechanical and other types of oscillators. Consider the waveguide structure of Fig. 1. The individual waveguides of this structure are slab waveguides (i.e. they are of infinite extent in the y coordinate that is perpendicular to the plane). Their fields can therefore be expressed in rectangular coordinates without coupling between the fields along the coordinate axes. This greatly simplifies the problem, because it allows us to find each of the field components as a solution to the scalar wave equation. The full mode structure in the region where the waveguides are in close proximity is considerably more complex, so we are compelled to search for approximations that will give us analytical solutions.

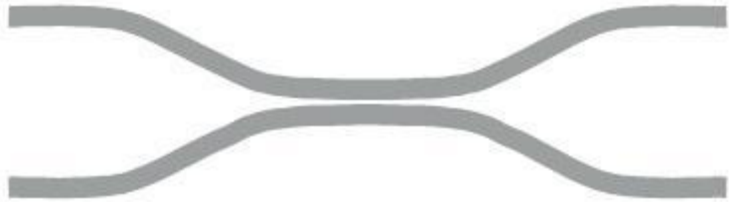

**Figure 1** Waveguide structure consisting of two waveguides that are brought into close proximity over a finite distance. In the area where the two waveguides are close, the modes of the waveguides will interact, which means that their propagation vectors will be shifted, and their modes modified.

Our starting point is the scalar wave equation:

$$\nabla^2 E_y(x,z) = \varepsilon\mu \frac{\partial^2 E_y(x,z)}{\partial t} \qquad (1)$$

that is valid in all the regions of the waveguide. The electric permittivity, or dielectric constant, and the magnetic permeability are both time invariant, so the solutions we are interested in are independent of z, i.e. they are of the form

$$E_{yi}(x,z) = \frac{1}{2} A_i . u_{yi}(x) e^{-j(\beta_i z - \omega t)z} + c.c \qquad (2)$$

If we now chose to first consider the fundamental TE mode of a symmetric waveguide, then the mode profile is given by

$$u_y(x,z) = \begin{cases} C \exp[-\gamma_c x] & x > 0 \\ C \left[\cos(\kappa_k x) - \frac{\gamma_c}{\kappa_k} \sin(\kappa_k x)\right] & -h < x < 0 \\ C \left[\cos(\kappa_k x) + \frac{\gamma_c}{\kappa_k} \sin(\kappa_k x)\right] \exp[-\gamma_s(x+h)] & x < -h \end{cases} \qquad (3)$$

where is C is determined by the requirement

$$\int_S u_{yn}(x) u_{ym}(x) \, dx = \frac{2\omega \mu_0}{\beta_n} \delta_{mn} \qquad (4)$$

As mentioned above, solving the wave equation on coupled waveguide structures is prohibitively hard and cannot be done analytically. Our approach to this problem is to consider the simpler structure, for which we can find the modes, and introduce the coupling as a perturbation of the polarization of the medium. To do that, we write constitutive relation for electric field as

$$\vec{D} = \varepsilon \vec{E} = \varepsilon_0 \vec{E} + \vec{P} = \varepsilon \vec{E} + \vec{P}_{pert} \qquad (5)$$

which means we can write the wave equation in the following way,

$$\nabla^2 E_y(x,z,t) = \varepsilon \mu \frac{\partial^2 E_y(x,z,t)}{\partial t} = \varepsilon \mu \frac{\partial^2 E_y(x,z)}{\partial t} + \mu \frac{\partial^2 P_{pert}(x,z,t)}{\partial t} \qquad (6)$$

First we set the perturbation to zero and find the modes of the unperturbed or uncoupled waveguide. The fields of the perturbed guide can be expressed in terms of these unperturbed modes

$$E_y(x) = \frac{1}{2} \sum_i A_i^+ u_i(x) . \exp[-j(\beta_i . z - \omega . t)] + c.c$$

$$+ \frac{1}{2} \sum_i A_i^- u_i(x) . \exp[-j(\beta_i . z + \omega . t)] + c.c$$

$$+ \frac{1}{2} \int_{k_0 n_s}^{k_0 n_f} A(\beta,z) . u_\beta(x) . \exp[-j(\beta . z + \omega . t)] d\beta + c.c \qquad (7)$$

where we have included both forward and backward traveling waves. If we carry the full field expansions, we haven't made any approximations, but the problem isn't simplified either. To make the problem feasible, we will assume that coupling to the radiation modes is negligible and therefore drop these modes from the expansion. Coupled-mode theory relies on this approximation, so negligible coupling to radiation modes can therefore be used as a criterion for when to apply coupled mode theory.

Now substitute the expanded solution back into the wave equation

$$\nabla^2 \{\sum_i A_i^+ u_i(x).\exp[-j(\beta_i.z - \omega.t)]$$

$$+ \frac{1}{2}\sum_i A_i^- u_i(x).\exp[j(\beta_i.z + \omega.t)] + c.c\}$$

$$= \mu \frac{\partial^2}{\partial t}\{\sum_i A_i^+ u_i(x).\exp[-j(\beta_i.z - \omega.t)]$$

$$+ \frac{1}{2}\sum_i A_i^- u_i(x).\exp[j(\beta_i.z + \omega.t)] + c.c\} + \mu \frac{\partial^2 P_{pert}(x,z,t)}{\partial t}$$

$$\Rightarrow \sum_i [A_i^+ \{-\beta_i^2 u_i(x) + \frac{\partial^2 u_i(x)}{\partial t} + \omega^2.\mu.\varepsilon.u_i(x)\}\exp[-j\beta_i z]$$

$$+ \{-2j\beta_i \frac{dA_i^+}{dz} + \frac{d^2 A_i^+}{dz^2}\} u_i(x) \exp[-j\beta_i z] + c.c] + \sum_i [A_i^- \{-\beta_i^2 u_i(x) + \frac{\partial^2 u_i(x)}{\partial t}$$

$$+ \omega^2.\mu.\varepsilon.u_i(x)\}\exp[-j\beta_i z] + \{-2j\beta_i \frac{dA_i^-}{dz} + \frac{d^2 A_i^-}{dz^2}\} u_i(x) \exp[j\beta_i z] + c.c]$$

$$= 2\exp[-j\omega t].\mu.\frac{\partial^2 P_{pert}(x,z,t)}{\partial t}$$

(8)

Notice that the first three terms of each of the two summations equals zero. The expression then simplifies to

$$\sum_i [\{-2j\beta_i \frac{dA_i^+}{dz} + \frac{d^2 A_i^+}{dz^2}\} u_i(x) \exp[-j\beta_i z]] + c.c + \sum_i [\{-2j\beta_i \frac{dA_i^-}{dz} + \frac{d^2 A_i^-}{dz^2}\} u_i(x) \exp[j\beta_i z]] + c.c$$

$$= 2\exp[-j\omega t].\mu.\frac{\partial^2 P_{pert}(x,z,t)}{\partial t}$$

(9)

We also assume slow variations of the amplitudes

$$\left|\frac{d^2 A_i}{dz^2}\right| \ll \beta_i \left|\frac{dA_i}{dz}\right|$$

(10)

So, we can write,

$$\exp[j\omega t]\frac{1}{2}\sum_i [\{-2j\beta_i \frac{dA_i^+}{dz}\} u_i(x) \exp[-j\beta_i z]] + c.c$$

$$+ \exp[j\omega t]\frac{1}{2}\sum_i [\{-2j\beta_i \frac{dA_i^-}{dz}\} u_i(x) \exp[j\beta_i z]] + c.c = \mu.\frac{\partial^2 P_{pert}(x,z,t)}{\partial t}$$

(11)

We multiply this equation with $u_i(x)$, and integrate over the cross section of the guide, and use mode orthogonality to arrive at our final result:

$$-\frac{dA_i^+}{dz}.\exp[j(\omega.t - \beta_i.z)] + \frac{dA_i^-}{dz}.\exp[j(\omega.t + \beta_i.z)] + c.c$$

$$= \frac{-j}{2\omega}\frac{\partial^2}{\partial t^2}\int_{-\infty}^{\infty} P_{pert}(x,z,t).u_i(x)dx$$

(12)

This equation can be used to treat a variety of waveguide structures with different types of interactions or coupling between guided modes. The exact form of the perturbation will depend on the waveguide structure at hand, but the general form of the coupled-mode equations will be the same.

## 2.2. Periodic Waveguides-Bragg Filters

We will now apply coupled mode theory to counter-propagating waves in a single mode waveguide with a periodic corrugation as shown in Fig. 2.

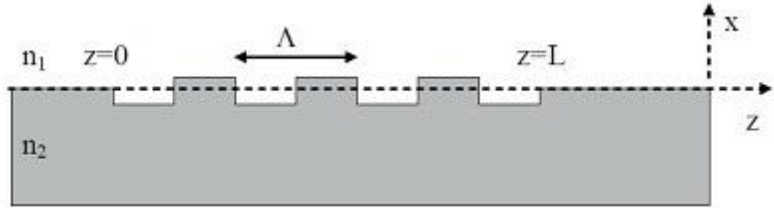

**Figure 2** Waveguide with periodic corrugation in one of the core-cladding interfaces. The waveguide is single mode, and we make the assumption that the only significant coupling is between counter propagating guided modes

The corrugation is scalar and we don't expect coupling between TE and TM modes, so in the following treatment we'll consider TE modes. We start by describing the field in the corrugated structure as a sum of the forward and backward propagating modes

$$E_y = A(z)u(x)\exp[j(\omega.t - \beta.z)] + B(z)u(x)\exp[j(\omega.t + \beta.z)] \quad (13)$$

where A and B are the amplitudes of the forward and backward propagating waves, and u(x) is the mode profile.

The perturbation in the corrugated region is

$$\vec{P}_{pert} = \Delta n(x,z)^2 \varepsilon_0 \vec{E} \quad (14)$$

We substitute the expression for the field into this expression to get

$$P_{pert} = \left(\frac{1}{2}\right)\Delta n(x)^2 \varepsilon_0 \{A(z)u(x)\exp[j(\omega.t - \beta.z)] + B(z)u(x)\exp[j(\omega.t + \beta.z)]\} \quad (15)$$

$$P_{pert} = \left(\frac{1}{2}\right)\Delta n(x)^2 \varepsilon_0 e^{j\omega.t} e^{-j\beta.z}\{A + Be^{-j2\beta.z}\}u(x) \quad (16)$$

Recall the fundamental coupled mode equation

$$-\frac{dA_i^+}{dz}.\exp[j(\omega.t - \beta_i.z)] + \frac{dA_i^-}{dz}.\exp[j(\omega.t + \beta_i.z)] + c.c$$

$$= \frac{-j}{2\omega}\frac{\partial^2}{\partial t^2}\int_{-\infty}^{\infty} P_{pert}(x,z,t).u_i(x)dx \quad (17)$$

which simplifies to

$$-\frac{dA}{dz} + \frac{dB}{dz}e^{j2\beta.z} = \frac{-j\omega.\varepsilon_0}{4}\{A + Be^{j2\beta.z}\}\int_{-\infty}^{\infty}\Delta n^2 u^2(x)\,dx \quad (18)$$

We will now assume that the corrugation has a square-wave shape as indicated in Fig. 4. The general conclusions are not dependent on the exact shape, so the following treatment, with appropriate adjustments, is valid also for non-square corrugations. The square-wave corrugations can be expressed as a series in the following form

$$\Delta n^2(x,z) = \Delta n^2 \sum_m C_m e^{j \cdot \frac{2m\pi \cdot z}{\Lambda}} \tag{19}$$

$$C_m = \begin{cases} -\dfrac{j}{m\pi} & m = odd \\ 0 & m = even \end{cases} \tag{20}$$

By comparing this expression to the above coupled mode equations, we realize that only modes that are close to phase matched will experience significant coupling. In other words, we need only keep terms of the same periodicity. In a range of wave vectors, the equations can be simplified to

$$\frac{dA}{dz} = \frac{j\omega.\varepsilon_0}{4} B e^{j2\beta \cdot z} C_m e^{-j \cdot \frac{2m\pi \cdot z}{\Lambda}} \int_{-\infty}^{\infty} \Delta n^2 u^2(x) dx \tag{21}$$

$$\frac{dB}{dz} = \frac{j\omega.\varepsilon_0}{4} A e^{-j2\beta \cdot z} C_m e^{j \cdot \frac{2m\pi \cdot z}{\Lambda}} \int_{-\infty}^{\infty} \Delta n^2 u^2(x) dx \tag{22}$$

$$\frac{dA}{dz} = K^* B e^{j2\Delta\beta \cdot z} \tag{23}$$

$$\frac{dB}{dz} = K A e^{-j2\Delta\beta \cdot z} \tag{24}$$

$$K = \frac{j\omega.\varepsilon_0}{4} C_m \int_{-\infty}^{\infty} \Delta n^2 u^2(x) dx \tag{25}$$

where

$$\Delta\beta = \beta - \frac{m\pi}{\Lambda} \tag{26}$$

Let us check energy conservation in the systems of equations we have found for modes in a Bragg grating. We start by deriving expression for the energies in the forward and backward propagating waves. Based on Eqns. 23 and 24 we can write

$$\frac{d}{dz}|A|^2 = \frac{d}{dx}(A.A^*) = A.\frac{dA^*}{dx} + A^*.\frac{dA}{dx} = A.B^*.K e^{-j2\Delta\beta \cdot z} + A^*.B.K^* e^{j2\Delta\beta \cdot z} \tag{27}$$

$$\frac{d}{dz}|B|^2 = \frac{d}{dx}(B.B^*) = B.\frac{dB^*}{dx} + B^*.\frac{dB}{dx} = B^*.A.K e^{-j2\Delta\beta \cdot z} + B.A^*.K^* e^{j2\Delta\beta \cdot z} \tag{28}$$

The difference between the rates of change in the forward-propagating and backward-propagating energy is then

$$\frac{d}{dz}|A|^2 - \frac{d}{dz}|B|^2 = A.B^*.K e^{-j2\Delta\beta \cdot z} + A^*.B.K^* e^{j2\Delta\beta \cdot z}$$
$$-B^*.A.K e^{-j2\Delta\beta \cdot z} + B.A^*.K^* e^{j2\Delta\beta \cdot z} = 0 \tag{29}$$

We see that the rate of change in forward-propagating energy is exactly balanced by the rate of change in backward-propagating energy, which is the correct result for loss-less, counter-propagating waves.

### 2.3. Modes of the Bragg Grating

The set of equations describing the modes of the Bragg Grating (Eqns. 23-26) can now be solved. Assuming that the forward propagating mode has an amplitude $A_0$ at z=0, and that the backward propagating wave is zero at z=L, we find

$$A = A_0 e^{j\Delta\beta z} \cdot \frac{-\Delta\beta \sinh[S(z-L)] + j\, S\cosh[S(z-L)]}{-\Delta\beta \sinh[SL] + j\, S\cosh[SL]} \quad (30)$$

$$A = A_0 e^{j\Delta\beta z} \cdot \left[\frac{-\Delta\beta \sinh[S(z-L)]}{-\Delta\beta \sinh[SL] + j\, S\cosh[SL]} + \frac{j\, S\cosh[S(z-L)]}{-\Delta\beta \sinh[SL] + j\, S\cosh[SL]}\right] \quad (31)$$

$$B = A_0 \cdot jK \cdot e^{-j\Delta\beta z} \cdot \frac{\sinh[S(z-L)]}{-\Delta\beta \sinh[SL] + j\, S\cosh[SL]} \quad (32)$$

where

$$S = \sqrt{K^2 - \Delta\beta^2} \quad (33)$$

When $\Delta\beta = 0$, this simplifies to

$$A = A_0 \cdot \frac{\cosh[K(z-L)]}{\cosh[KL]} \quad (34)$$

$$B = A_0 \cdot \frac{\sinh[K(z-L)]}{\cosh[KL]} \quad (35)$$

### 2.4. Bragg Filters

The expressions we have found for the field amplitudes in the periodically corrugated waveguide allow us to calculate the reflection and transmission spectra of the Bragg grating. For example, the field reflection is simply the ratio of the forward propagating and backward propagating wave at the input to the Bragg section:

$$r = \frac{B(0)}{A(0)} = \frac{A_0 \cdot jK \frac{\sinh[SL]}{-\Delta\beta \sinh[SL] + j\, S\cosh[SL]}}{A_0 \cdot \frac{-\Delta\beta \sinh[SL] + j\, S\cosh[SL]}{-\Delta\beta \sinh[SL] + j\, S\cosh[SL]}} \Rightarrow$$

$$r = j \cdot K \frac{\sinh[SL]}{-\Delta\beta \sinh[SL] + j\, S\cosh[SL]} \quad (36)$$

### 3. Results and Discussion

Using Eq (30) and Eq (31), forward and backward waves are calculated as a function of grating length for different input wavelengths. Fig 3 shows the profile for forward waves, whereas Fig 4 is plotted for backward waves. From Fig 3 it can be seen that magnitude of wave is maximum when input wavelength matches with Bragg wavelength, which is the resonance condition. This variation is significant for small grating length. If input wavelength is greater or less than that of the corresponding Bragg value, then the magnitude decreases. For backward wave, the value is lowest for resonance condition, which eventually satisfies the energy conservation principle. The difference in magnitude increases with higher value of grating length, as may be observed from Fig 4. The rate of decrease of forward and backward waves is high for smaller grating length.

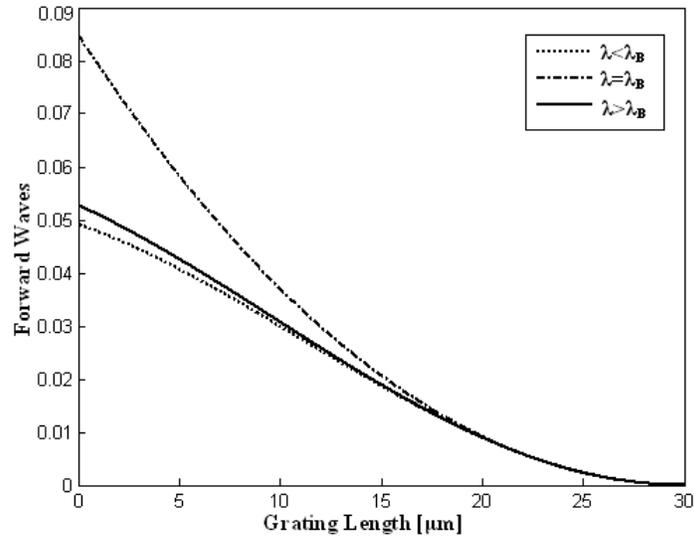

**Figure 3** Forward wave profiles with grating length for different input wavelengths

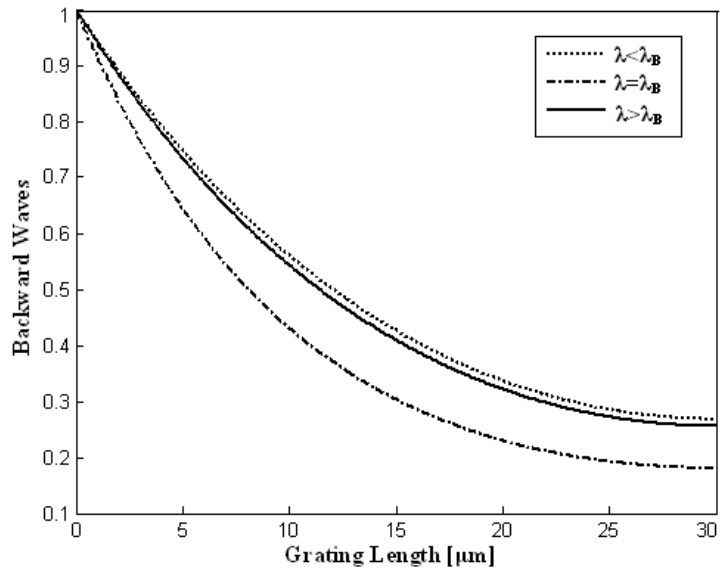

**Figure 4** Microscopic view of $Al_XGa_{1-X}N$ fabrication

Fig 5 exhibits reflectivity profile with wavelength for different grating lengths under weak coupling condition, calculated using Eq. (36). From the plot, it may be observed that reflectivity becomes higher for larger value of grating length. At resonance condition, due to the formation of photonic bandgap, most part of the light is reflected, and smaller portion is transmitted. Outside of this bandgap, reflectivity closes to zero, as we move away from Bragg wavelength. This phenomenon is also observed for strong coupling condition, and reflectivity reaches to unity for large grating length. This is plotted in Fig 6. A comparative study between these results indicates the fact that photonic bandgap is smaller for weak coupling condition, whereas it is wider for strong coupling condition. Also magnitude of reflectivity increases with increase of coupling coefficient.

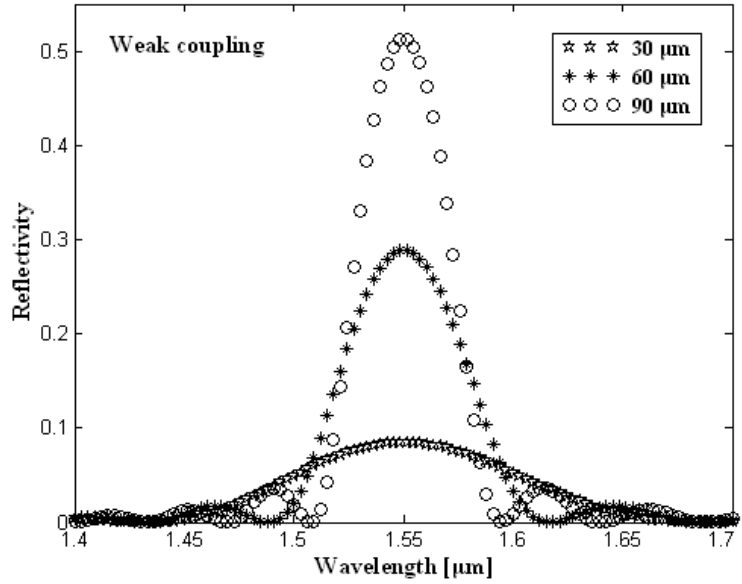

**Figure 5** Reflectivity with wavelength for different grating lengths under weak coupling conditions.

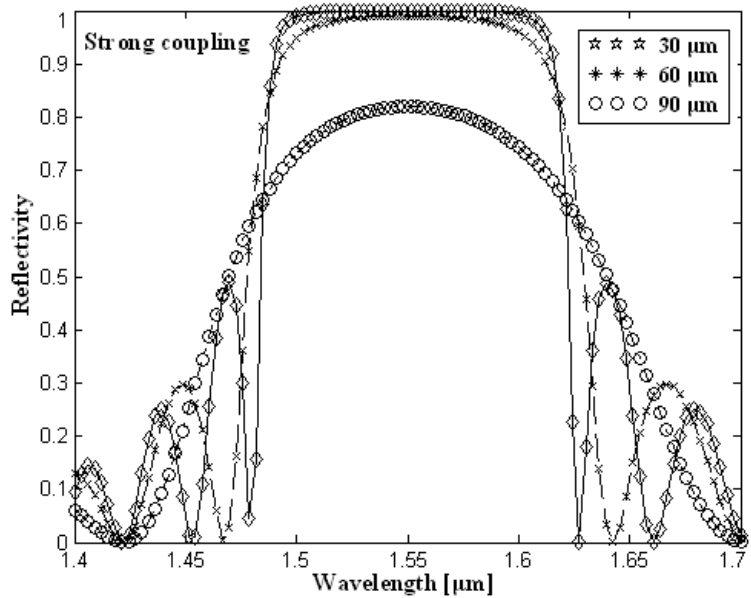

**Figure 6** Reflectivity with wavelength for different grating lengths under strong coupling conditions

Fig 7 shows the reflectivity profile with input wavelength for different coupling conditions and smaller grating lengths, whereas the same is plotted in Fig 8 for wider grating structure. If grating length is made higher, larger photonic bandgap is formed for strong coupling condition, and almost hardly any transmission of electromagnetic wave takes place. This complete reflection condition was absent if coupling is made relatively weak. Similarly, for smaller value of grating length, percentage of transmission and reflection, i.e., extent of the photonic bandgap depends on coupling effect centered on Bragg wavelength. It may also be noted down that for the present structure, 50% reflection can be obtained for 30 μm grating length when coupling coefficient is 0.03μm$^{-1}$, which is the case of strong coupling. Similarly, almost no reflection is observed for the same grating length when coupling coefficient is 0.01 μm$^{-1}$, which is the case of weak coupling. For 90 μm grating length, spectral with becomes highest for 0.05 μm$^{-1}$ coupling coefficient, and peak value becomes 50% when κ becomes 0.01 μm$^{-1}$

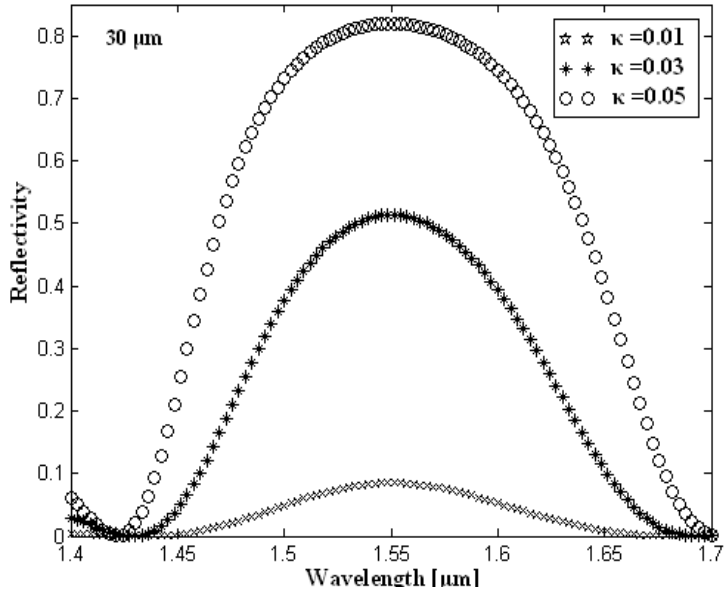

**Figure 7** Reflectivity with wavelength for different coupling coefficients with 30 μm grating length

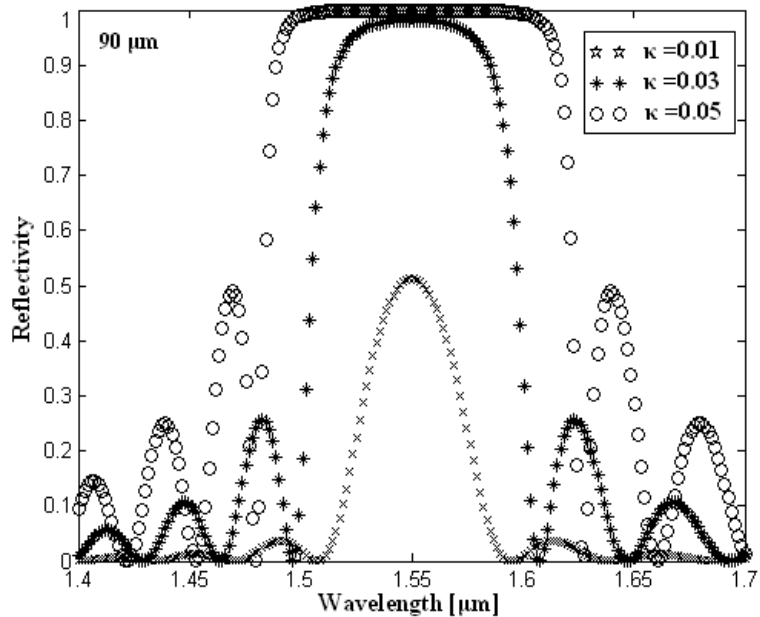

**Figure 8** Reflectivity with wavelength for different coupling coefficients with 90 μm grating length

## 4. Conclusion

From the analysis, it can be concluded that reflectivity, and magnitude of forward and backward travelling waves, depends on extent of grating along the propagation direction of electromagnetic wave, as well as coupling between forward and backward waves. Spectral width of photonic bandgap can be measured from reflectivity profiles, which is important in designing bandpass filters at optical communication spectra using Bragg reflectors.